\documentclass[aip,amsmath,amssymb,reprint]{revtex4-1}

\usepackage{graphicx}
\usepackage{dcolumn}
\usepackage{bm}
\usepackage[normalem]{ulem}
\usepackage[utf8]{inputenc}
\usepackage[T1]{fontenc}
\usepackage{mathptmx}
\usepackage{color} 
\usepackage[caption=false]{subfig}

\begin{document}

\title{Implementation of Perdew-Zunger self-interaction correction in real space using Fermi-L\"owdin orbitals}

\author{Carlos M. Diaz}
\email{cmdiaz6@miners.utep.edu}
\affiliation{Department of Physics, University of Texas at El Paso, El Paso, Texas 79968, USA\\}

\author{Phanish Suryanarayana}
\author{Qimen Xu}
\affiliation{College of Engineering, Georgia Institute of Technology, Atlanta, GA 30332, USA\\}

\author{Tunna Baruah}
 \affiliation{Department of Physics, University of Texas at El Paso, El Paso, Texas 79968, USA\\}

\author{John Pask}
\affiliation{Physics Division, Lawrence Livermore National Laboratory, Livermore, CA 94550, USA\\}

\author{Rajendra Zope}
 \affiliation{Department of Physics, University of Texas at El Paso, El Paso, Texas 79968, USA\\}

\date{\today}

\begin{abstract}
Most widely used density functional approximations suffer from  self-interaction (SI) error, 
which can be corrected using the Perdew-Zunger (PZ) self-interaction correction (SIC). 
We implement the recently proposed size-extensive formulation of 
PZ-SIC using Fermi-L\"owdin Orbitals (FLOs) in real space, which is amenable 
to   systematic convergence and large-scale parallelization. 
We verify the new formulation
within the generalized Slater scheme
by computing atomization energies and ionization potentials of selected molecules
and comparing to those obtained by existing FLOSIC implementations in Gaussian based codes. 
The results show 
good agreement between the two formulations, with new real-space results somewhat closer to experiment on average for the systems considered. 
We also obtain the ionization potentials and atomization energies 
by scaling down the Slater statistical average of SIC potentials. 
The results show that scaling down the average SIC potential improves both atomization energies and ionization potentials, bringing them closer to experiment.
Finally, we verify the present formulation by calculating the barrier heights of chemical reactions in the BH6 dataset, where significant improvements are obtained relative to Gaussian based FLOSIC results.
\end{abstract}

\maketitle

\section{Introduction}

The Kohn-Sham (KS) formulation of the density functional theory (DFT) has
emerged as a standard method for electronic structure calculations of atoms, molecules, 
and solids.\cite{jones2015density} 
It is among the most widely used quantum mechanical methods for the
study of electronic properties of solids. 
The many-electron effects in the KS theory are incorporated into the exchange-correlation functional, the exact form of which is currently unknown.  
Practical applications of KS-DFT therefore employ  an approximation to the exchange-correlation functional, the accuracy
of which determines the fidelity of the simulation. Since there is no systematic
way to construct the exchange-correlation functional, a large number of approximations have been
been proposed.\cite{perdew2001jacob,MARQUES20122272}  Both exchange and
correlation are generally approximated together to obtain systematic error
cancellation.  One way of constructing functionals, pursued by Perdew and
coworkers, is by considering the known physical properties of the exact
exchange-correlation functional and  other exact 
properties.\cite{perdew1982density,perdew2001jacob,tsuneda2000parameter} Such functionals,
classified as non-empirical, 
typically show greater predictive capability than functionals empirically fitted to
databases of various chemical properties.

Perdew and Schmidt\cite{perdew2001jacob} have classified exchange-correlation
functionals using an analogy to Jacob's ladder wherein more sophisticated density
functional approximations (DFA) correspond to higher rungs on the  ladder. The
celebrated local spin density approximation (LSDA) functional,\cite{PhysRev.140.A1133,Perdew1981}
which depends only on the electron
density, belongs to the first rung of the ladder. This functional has been used
for decades to study electronic structure related properties in solid state
physics. Semi-local functionals such as the generalized gradient
approximation (GGA) depend on the electron density and its gradients, while
meta-GGA functionals additionally depend on the Laplacian of the density or kinetic-energy
density. These functionals belong to the second and third rungs, respectively.
The higher rungs' functionals include a certain percent of non-local Hartree-Fock
exchange along with DFAs. \cite{perdew2001jacob,MARQUES20122272}
Acceptable accuracy of the available functionals along
with efficient implementation of KS formulations in several easy-to-use codes 
have resulted in explosive growth of KS-DFT applications. 

The LSDA and
semi-local functionals often describe equilibrium properties sufficiently
accurately but fail to provide accurate results for reaction barrier heights in
chemical reactions where the bonds between atoms are 
stretched.\cite{doi:10.1063/1.1370527} In fact, these functionals
also fail to describe one electron systems accurately, which has been attributed
to the self-interaction error (SIE) of these functionals. \cite{doi:10.1063/1.1370527} Specifically, in the case of a one electron
system, the exact exchange-correlation energy cancels the Coulomb energy. However, this cancellation is incomplete in
approximate functionals, and residual
self-interaction remains. A number of approaches have been proposed to remove 
the SIEs.\cite{lindgren1971statistical,perdew1979orbital,PhysRevA.15.2135,Perdew1981,gunnarsson1981self,lundin2001novel,
doi:10.1063/1.2176608, doi:10.1063/5.0004738,doi:10.1063/1.5129533}
In this work, we focus on the most widely used self-interaction-correction, developed by Perdew and Zunger (PZ-SIC)\cite{Perdew1981}, 
which has been used to study atoms, molecules, and solids.\cite{Heaton1983,doi:10.1063/1.446959,doi:10.1063/1.448266,
     doi:10.1063/1.481421, doi:10.1063/1.1327269, doi:10.1063/1.1370527, harbola1996theoretical, doi:10.1063/1.1468640, doi:10.1021/jp014184v,PhysRevA.55.1765,doi:10.1080/00268970110111788, Polo2003, doi:10.1063/1.1630017, B311840A,doi:10.1063/1.1794633, doi:10.1063/1.1897378, doi:10.1063/1.2176608, zope1999atomic, doi:10.1063/1.2204599,doi:10.1002/jcc.10279,PhysRevA.45.101, PhysRevA.46.5453,lundin2001novel, PhysRevA.47.165,doi:10.1021/acs.jctc.6b00347,csonka1998inclusion,petit2014phase,kummel2008orbital,schmidt2014one,kao2017role,schwalbe2018fermi,jonsson2007accurate,rieger1995self,temmerman1999implementation,daene2009self,szotek1991self,messud2008time,messud2008improved,doi:10.1063/1.1926277,korzdorfer2008electrical,Korzdorfer2008,ciofini2005self,PhysRevA.50.2191,
    doi:10.1063/1.5125205,C9CP06106A,doi:10.1002/jcc.25767,doi:10.1021/acs.jctc.8b00344,doi:10.1063/1.4947042,schwalbe2019pyflosic,
    doi:10.1063/1.4996498, doi:10.1063/1.5050809, doi:10.1021/acs.jpca.8b09940,Jackson_2019, Sharkas11283}

\subsection{Perdew-Zunger Self-interaction-correction (PZ-SIC)}
        In 1981, Perdew-Zunger (PZ) provided a
definition of one-electron SIE and outlined a procedure to eliminate this error
from the DFAs.\cite{Perdew1981} In the PZ approach, the SIE is removed on an orbital by orbital
basis according to the following equation:

\begin{equation} \label{eq:PZSIC}
 E^{SIC}=-\sum_{i\sigma}^{N_{occ}} \left ( U[\rho_{i\sigma} ]
 + E_{xc}^{DFA} [\rho_{i\sigma},0] \right ).
\end{equation}
Here, $U[\rho_{i\sigma}]$ and $E_{xc}^{DFA} [\rho_{i\sigma},0 ]$ 
are the Coulomb and exchange-correlation energy of the $i^{th}$ occupied orbital,
$\sigma$ is the spin index and $N_{occ}$ is the number of occupied orbitals. 
The orbital electron density, $\rho_{i\sigma}=|\psi_{i\sigma}|^2$, where the $\psi_{i\sigma}$ are the KS orbitals. 
The above corrections make the DFA exact for any one-electron density. 
The correction should vanish for the exact functional. The SIC method of Perdew-Zunger
satisfies this criterion.
The total energy with the PZ-SIC method is given by 
\begin{equation}
E =E^{KS}+E^{SIC},
\end{equation}
where, 
\begin{multline}
    E^{KS}=\sum_{i\sigma} f_{i\sigma} \langle \psi_{i\sigma}|-\frac{\nabla^2}{2}|\psi_{i\sigma} \rangle 
    + \int d^3r \, \rho(\vec{r})v_{ext}(\vec{r})  \\
    +\frac{1}{2}\iint d^3r \, d^3r' \frac{\rho(\vec{r})\rho(\vec{r}')}{|\vec{r}-\vec{r}'|}
    +E_{xc}[\rho_{\uparrow},\rho_{\downarrow}].
\end{multline}
Here, $v_{ext}$ is the external potential. The electron density, $\rho = \rho_{\uparrow} + \rho_{\downarrow}
=  \sum_{\sigma} \rho_{\sigma} = \sum_{i,\sigma} f_{i\sigma} \vert \psi_{i\sigma}\vert^2 $.
$f_{i\sigma}$ is the occupation of the $\psi_{i\sigma}$ orbital.

The variational minimization of the above energy functional results in a set of orbital dependent 
PZ-SIC equations: 
\begin{equation} \label{eq:KS-PZ}
\left[-\frac{\nabla^2}{2}  + v_{eff}^{i\sigma}(\vec{r})\right] \psi_{i\sigma}
 = \hat{H}^{PZ-SIC}_{i\sigma}~ \psi_{i\sigma}
  = \sum_{j=1,\sigma}^N \lambda_{ji}^\sigma\psi_{j\sigma}.
\end{equation}
Here,
\begin{eqnarray}
\label{eq:Veff}
v_{eff}^{i\sigma}(\vec{r}) & =  & v_{ext}  (\vec{r})
+ 
\int d^3r'\, \frac{\rho(\vec{r}')} {\vert \vec{r}-\vec{r}'\vert}  +  
v_{xc}^{\sigma} (\vec{r}) \nonumber \\
& - &
 \left \{ \int d^3r' \,
 \frac{\rho_{i\sigma} (\vec{r}')} {\vert \vec{r}-\vec{r}'\vert}  +  
v_{xc}^{i\sigma} (\vec{r})
 \right \}.
\end{eqnarray} 
The first term  $ v_{ext}  (\vec{r})$ in above equation 
is the external potential, the second term is the Coulomb 
potential due to the electrons, and 
$v_{xc}$ is the exchange-correlation potential (of DFA). The last two terms enclosed in the curly bracket 
correspond to the orbital SIC potential $V_{i\sigma}^{SIC}$, 
composed of the self-Coulomb and self-exchange-correlation potential. 
Note that, unlike the standard KS equations, the effective potential in  Eq.~(\ref{eq:KS-PZ})  is orbital dependent. 
The matrix of Langrage multiplers 
$\lambda_{ij} =  \langle  \psi_{i\sigma} \vert  \hat{H}^{PZ-SIC}_{i\sigma} \vert  \psi_{j\sigma} \rangle $
is introduced in Eq. \ref{eq:KS-PZ} to ensure the orthogonality of occupied $N$ orbitals of spin $\sigma$.
At the minimum of the PZ-SIC energy, the non-Hermitian Lagrange multiplier matrix becomes Hermitian.
Thus, the orbitals minimizing the Hamiltonian in Eq. \ref{eq:KS-PZ} satisfy Pederson's localization 
equations\cite{doi:10.1063/1.448266,doi:10.1063/1.446959} $\lambda_{ji}^\sigma = \lambda_{ij}^\sigma .$
Some implementations\cite{doi:10.1063/1.1794633}
of PZ-SIC method enforce  satisfaction of  {\it localization equations}, a method which results
in additional computational effort. 
The orthogonal orbitals that satisfy the localization equations are localized orbitals. 
Various localized orbitals such as Foster-Boys\cite{Foster1960} and Pipek-Mezey\cite{Pipek1989}, for example, have been 
used in solving the PZ-SIC scheme. \cite{doi:10.1063/1.481421,doi:10.1063/1.1370527,doi:10.1063/1.1468640}
These are related to Kohn-Sham orbitals by 
unitary transformation. In 2014, Pederson and coworkers\cite{Pederson2014} used Fermi-L\"owdin orbitals\cite{Leonard1982,Luken1982,Luken1984}
in the PZ-SIC scheme. Below we describe in brief the details of that approach.

\subsection{Fermi-L\"owdin orbitals SIC (FLOSIC)}
 Recently, Pederson, Ruzsinszky, and Perdew \cite{Pederson2014} introduced the unitarily invariant 
     implementation of PZ-SIC using Fermi-L\"owdin orbitals (FLOSIC). \cite{Luken1982,Luken1984}
FLOSIC makes use of localized Fermi orbitals (FOs) $F_{i\sigma}$ which are defined by the transformation of KS orbitals as
\begin{equation}
    F_{i\sigma}(\vec{r})=\frac{\sum_\alpha \psi_{\alpha\sigma}^{*}(\vec{a}_{i\sigma})\psi_{\alpha\sigma}(\vec{r})}{ \sqrt{\sum_\alpha |\psi_{\alpha\sigma}(\vec{a}_{i\sigma})|^2}}.
\end{equation}
Here, $\vec{a}_{i\sigma}$ are points in space called Fermi-orbital descriptors (FODs). Neglecting the spin index,
the above equation can be rewritten as
\begin{equation}
    F_i(\vec{r})=\sum_{\alpha}^{N_{occ}}F_{i\alpha}\psi_{\alpha}=\frac{\rho(\vec{a}_i,\vec{r})}{\sqrt{\rho(\vec{a}_i)}},
\end{equation}
where the transformation matrix $F_{i\alpha}$ is defined as
\begin{equation} \label{eq:Tij}
    F_{i\alpha}=\frac{\psi_{\alpha}^*(\vec{a}_i )}{\sqrt{\rho(\vec{a}_i)}}.
\end{equation}
The FOs are normalized but not orthogonal, so they are orthogonalized using L\"owdin orthogonalization method 
to generate the Fermi-L\"owdin orbitals (FLOs) $\phi_{i\sigma}$. The FLOs depend on 
the FOD positions. The change of FOD positions thereby alters the total energy.  The optimal 
FOD positions are obtained by minimizing the self-consistent SIC energy using the FOD forces\cite{Pederson2015,PEDERSON2015153}
as follows
\begin{eqnarray} \label{eq:fodforce}
    \frac{dE^{SIC}}{da_m}=\sum_{kl}\epsilon^k_{kl}
    \left \{ \left \langle \frac{d\phi_k}{da_m}|\phi_l \right \rangle 
    - \left \langle \frac{d\phi_l}{da_m}|\phi_k \right \rangle   \right \} .
\end{eqnarray}
The derivatives 
$\vert (d\phi_k)/(da_m )\rangle$ require evaluating
$\vert (dF_k)/(da_m )\rangle $
which can be accomplished using following relations,
\begin{equation}
    \nabla_{a_i}F_i(\vec{r})=\sum_\alpha\{\nabla_{\vec{a_i}}F_{i\alpha}\} \psi_a(\vec{r})\}
\end{equation}
\begin{equation}
    \nabla_{a_i}F_{i\alpha}=F_{i\alpha} \Big\{\frac{\nabla_{a_i}\psi_\alpha(\vec{a}_i)}{\psi_\alpha(\vec{a}_i)} - \frac{\nabla_{a_i}\rho(\vec{a}_i)}{2\rho(\vec{a}_i)} \Big\}
\end{equation}

The FLOSIC approach described above, unlike the traditional PZ-SIC implementations, requires optimizing the $N$ positions or $3N$ parameters of FODs and has formally lower cost than the traditional PZ-SIC implementation which requires obtaining $N^2$ 
coefficients to satisfy the localization equations.  The FLOSIC method has been 
implemented in Gaussian-orbital-based packages and has been successfully applied 
to study a wide array of electronic properties.\cite{doi:10.1063/1.5125205, C9CP06106A, doi:10.1002/jcc.25767, doi:10.1063/1.5050809, doi:10.1063/1.4996498,Jackson_2019, PhysRevA.100.012505,
doi:10.1021/acs.jpca.8b09940,
doi:10.1063/1.5087065,
doi:10.1063/1.5129533,
doi:10.1021/acs.jctc.8b00344,
Sharkas11283,
Jackson_2019,
doi:10.1063/1.5120532,
SingHam,
doi:10.1063/5.0004738,
schwalbe2018fermi,
schwalbe2019pyflosic}

In this work, we present a formulation and implementation of the FLOSIC method in real space. 
The real-space formulation enables rigorous, systematic convergence for all atomic species and configurations as well as large-scale parallelism to reach larger length and time scales than previously accessible at this level of theory.
While the PZ-SIC method using Kohn-Sham orbitals has been implemented in real space previously \cite{messud2008improved, messud2008time, kummel2008orbital, doi:10.1063/1.4865942, Korzdorfer2008, korzdorfer2008electrical, schmidt2014one}
we present the first implementation of the FLOSIC method, to our knowledge.
In Sec.~\ref{sec:FD_implementation}, we present 
the details of the formulation and implementation.
To demonstrate and verify the formulation and implementation, we compute the atomization energies, barrier heights,
and ionization potentials of selected molecules and compare to previous results obtained by standard Gaussian based codes 
and experiment. Computational details and results are presented in Secs.~\ref{sec:computational_details} and \ref{sec:results}, respectively.

\section{Real-space formulation and implementation}\label{sec:FD_implementation}
\subsection{M-SPARC code base}
To facilitate rapid implementation and testing, we formulate and implement the self-interaction correction in real space in the M-SPARC prototype code, \cite{xu2020m} a serial implementation of the massively parallel SPARC real-space electronic structure code, \cite{ghosh2017sparc2,ghosh2017sparc1,xu2020sparc} sharing the same structure, algorithms, input, and output.

M-SPARC is an open-source software package that can perform Kohn-Sham DFT calculations for isolated systems such as molecules as well as extended systems such as crystals and surfaces. M-SPARC employs the pseudopotential approximation \cite{Martin2004} to facilitate the efficient solution of the Kohn-Sham equations for all elements in the periodic table. In addition, it  employs a local real-space formulation for the electrostatics, \cite{pask2005elstat,Suryanarayana2014524,ghosh2014higher} wherein the electrostatic potential --- sum of ionic and Hartree potentials --- is obtained via the solution of a Poisson equation. In this framework, M-SPARC performs a uniform real-space discretization of the Kohn-Sham and Poisson equations, using a high-order centered finite-difference approximation for differential operators and the trapezoidal rule for integral operators. 

M-SPARC employs the self-consistent field (SCF) method \cite{Martin2004} to solve for the electronic ground state. The superposition of isolated-atom electron densities is used as the initial guess for the first SCF iteration in the simulation, whereas for every subsequent atomic configuration encountered, extrapolation based on  previous configurations' solutions is employed. \cite{alfe1999ab} The SCF method is accelerated using the  restarted variant \cite{pratapa2015restarted} of Periodic Pulay mixing  \cite{banerjee2016periodic} with real-space preconditioning. \cite{kumar2019preconditioning} For spin polarized calculations, mixing is performed simultaneously on both components. 

In every SCF iteration, M-SPARC performs a partial diagonalization of the linear eigenproblem using the CheFSI method, \cite{zhou2006self,zhou2006parallel} with multiple Chebyshev filtering steps performed in the first SCF iteration of the simulation \cite{zhou2014chebyshev}. 
While performing the Hamiltonian-matrix products, the Kronecker product formulation for the Laplacian is used, \cite{sharma2018real} and the remaining terms are handled in a matrix-free fashion
with zero-Dirichlet or Bloch-periodic boundary conditions  prescribed on the orbitals along directions in which the system is finite or extended, respectively. The electrostatic potential is determined by using the AAR linear solver \cite{pratapa2016anderson,suryanarayana2019alternating} for solving the Poisson equation. Laplacian-vector products are performed 
using the Kronecker product formulation, \cite{sharma2018real}
with Dirichlet/periodic boundary conditions prescribed along directions in which the system is finite or extended, respectively. In particular, Dirichlet values are determined using a multipole expansion for isolated systems and a dipole correction for surfaces and nanowires. \cite{burdick2003parallel,natan2008real}

M-SPARC thus provides a natural framework to implement the FLOSIC self-interaction correction in real space, for both isolated and periodic systems. Its extension to SPARC should be relatively straightforward, providing large-scale parallelism for performing such calculations.

\subsection {Interpolation Scheme: Transformation matrix and FOD forces}
     In order to construct the transformation matrix $F_{ij}$ (Eq.~(\ref{eq:Tij})) 
     the wavefunction $\psi(\vec{a}_i)$ needs to be evaluated at the FOD position
     $\vec{a}_i$, which can fall between grid points. In such case,
     values between grid points are obtained by cubic interpolation of values on the grid. 
     The density $\rho (\vec{a}_i)$ is then constructed as 
     $ \rho (\vec{a}_i) = \sum_j^N f_j \vert \psi_j (\vec{a}_i) \vert^2 $, where $f_j$ is the occupancy of the $j$-th state.
    Currently all calculations use a fixed integer occupation.
    Calculating the forces on FODs according to Eq.~(\ref{eq:fodforce}) also requires 
    the gradients $\nabla\psi(\vec{a}_i)$ at the FOD positions $\vec{a}_i$. We first obtain the gradient at each 
    grid point using  the 1st-order central-difference approximation for all central 
    points, and forward- and backwards-difference at the edges. 
    The gradients 
    $\nabla\psi(\vec{a}_i)$ are then obtained by cubic interpolation as for the transformation matrix, $F$.
    Proceeding in this way, forces and energy minima were found to coincide well. 
    For Li$_2$, analytic FOD forces differed by $2.6\times 10^{-6}$
    $E_h$/Bohr at equilibrium when compared against finite-difference force calculations using a spacing of $10^{-2}$ Bohr.

\subsection{Self-consistency} \label{sec:SC}
   The development of full self-consistency in FLOSIC was shown to lower energies
   for certain systems.\cite{Yang2017} In their work, Yang \textit{et al.} employed an iterative approach to solve
   Eq.~(\ref{eq:KS-PZ}) using Jacobi rotations (SC$_{Jacobi}$). 

In this approach, an approximate 
    Hamiltonian matrix
is first constructed as
\begin{equation}
    \tilde{H}_{mn\sigma}=\langle \phi_{m\sigma} | \hat{H}_\sigma^{KS} + V_{m\sigma}^{SIC} | \phi_{n\sigma} \rangle
\end{equation}
where $\hat{H}_\sigma^{KS}$ is the traditional KS Hamiltonian.
The FLOs and the unoccupied virtual orbitals are made orthogonal through pairwise Jacobi rotations
which are carried out iteratively until the matrix elements for $\tilde{H}$ between the occupied $\phi_i$ and virtual orbitals $\phi_j$ vanishes.
   
   An alternative approach has since been 
   introduced which uses the Unified Hamiltonian described by Heaton, Harrison, 
   and Lin \cite{Heaton1983} (SC$_{UH}$). In both cases, operations are performed on the full Hamiltonian 
   of size $N_b^2$, where $N_b$ is the number of basis functions in the calculation.
   In the finite-difference methodology, the Hamiltonian is of order $N_{grid}^2$, where $N_{grid}$ is the 
   total number of grid points used in the calculation as determined by the domain
   size and grid spacing. The size of the real-space Hamiltonian would quickly make 
   these approaches prohibitively expensive. 
     One approach to address this would be to employ discrete discontinuous basis 
 projection (DDBP)\cite{xu2018discrete}
to reduce Hamiltonian and orbital dimensions to a few tens per atom. 

\begin{figure}
    \centering
    \includegraphics[height=0.4\paperheight]{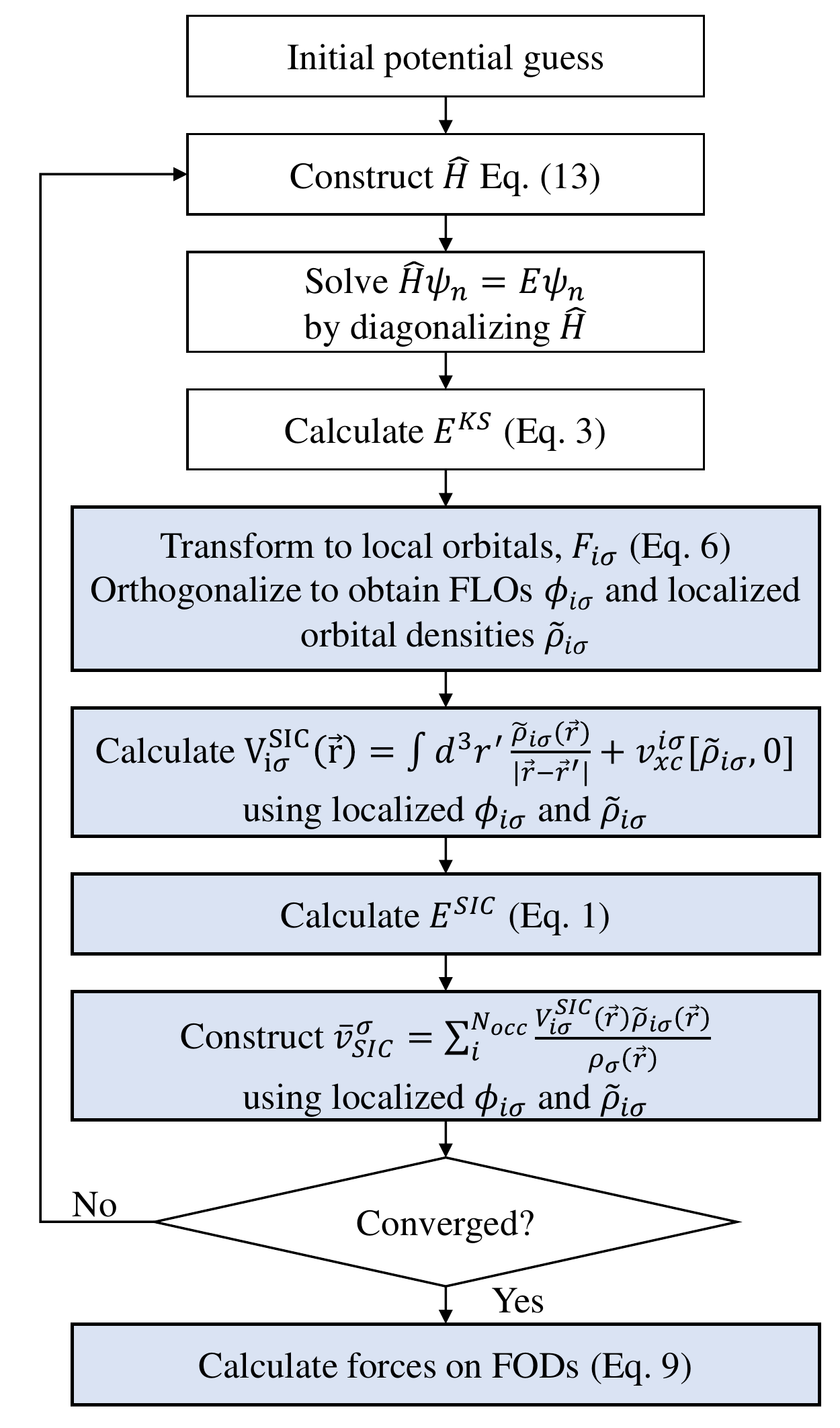}
    \caption{Self-consistent FLOSIC procedure using the averaged SIC potential scheme for a given set of FODs. Steps making use of localized FLOs are indicated by color and greater width.}
    \label{fig:scf}
\end{figure}

In the present work 
we use a simplified single Hamiltonian which is obtained by statistical 
average of the orbital dependent SIC potentials\cite{Slater1951} (SC$_{AvgSIC}$). 
This eliminates the dependence on the full $N_b^2$ matrix elements.
Thus the Hamiltonian used in the present work is given by
\begin{eqnarray}
 \hat{H} = -\frac{\nabla^2}{2}  
 + v_{ext}  (\vec{r})
+ 
\int d^3r'\, \frac{\rho(\vec{r}')} {\vert \vec{r}-\vec{r}'\vert}  +  
v_{xc}^{\sigma} (\vec{r}) \nonumber \\
 + 
      \frac{\sum_i^{N_{occ}}~ V_{i\sigma}^{SIC} (\vec{r})~ \tilde{ \rho}_{i\sigma}(\vec{r}) 
              }{\rho_{\sigma}(\vec{r})} .
\end{eqnarray}
    Here, $ \tilde{\rho}_{i\sigma}(\vec{r})= \vert \phi_{i\sigma}(\vec{r}) \vert^2 $ is the FLO density, 
    and $ \rho_{\sigma}(\vec{r})$  is the spin
    electron density. 
     Figure \ref{fig:scf} illustrates the SCF procedure utilizing this approach.
     This approach resembles the traditional Slater average method but differs
     in two key respects. Firstly, the averaged potentials 
     contain both the exchange and correlation components. Secondly, 
     the average potential is constructed from orbital dependent potentials
     that are determined using the localized Fermi-L\"owdin orbital densities and not 
     the canonical (delocalized) Kohn-Sham orbital densities. Similarly, the 
     weights $\tilde{\rho}_i$ used in averaging are also FLO densities. 
     Such an approach using non-KS orbitals 
     has been referred to by Messud and coworkers as a {\it generalized Slater} 
     scheme.\cite{messud2008improved,messud2011generalized} 
     The difference from previous works 
     is the use of Fermi-L\"owdin orbitals 
     in the present work.
     We note that the average Slater potential is a further approximation of the 
      Krieger-Li-Iafrate approximation\cite{PhysRevA.46.5453} to the optimized effective potential (OEP)  where
      orbital-dependant shifts are ignored.
     Messud {\it et al.}\cite{messud2011generalized} have noted that the generalized Slater
     scheme provides a significant improvement compared to the traditional SIC-Slater
     and Krieger-Li-Iafrate formalisms using canonical 
     Kohn-Sham orbitals. 
     A number of studies have emphasized the need for localized orbitals 
     in PZSIC calculations on molecules and 
     solids\cite{doi:10.1063/1.446959,doi:10.1063/1.448266,Pederson_5scientific,doi:10.1063/1.481421,doi:10.1063/1.1370527,doi:10.1063/1.1468640,kummel2008orbital}
     since, as mentioned earlier, canonical orbitals give vanishing correction for extended 
     systems.
An added benefit of having a single effective potential is that the SI-corrected unoccupied orbitals 
are available in this approximation.

We have also experimented with scaling the average SIC potential; 
         J\'onsson and coworkers have found such a scaled SIC to be more accurate in some cases when used 
         with GGA functionals.\cite{Klupfel2012}

    \begin{figure}
        \centering
        \subfloat[N-LSDA]{\label{test1}
        \includegraphics[width=0.3\linewidth]{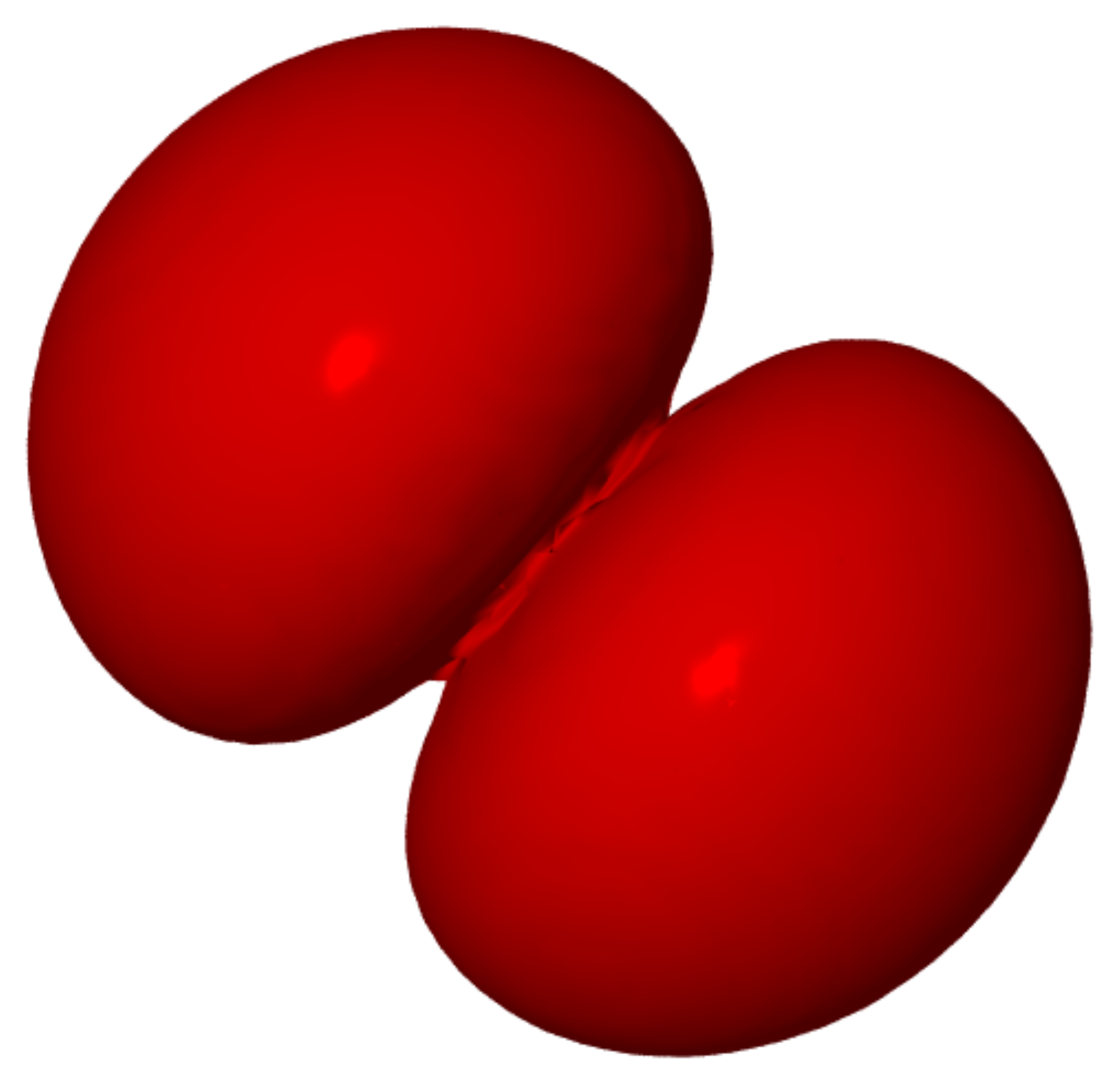}
        }
        \subfloat[N-SC$_{Jacobi}$]{
        \includegraphics[width=0.3\linewidth]{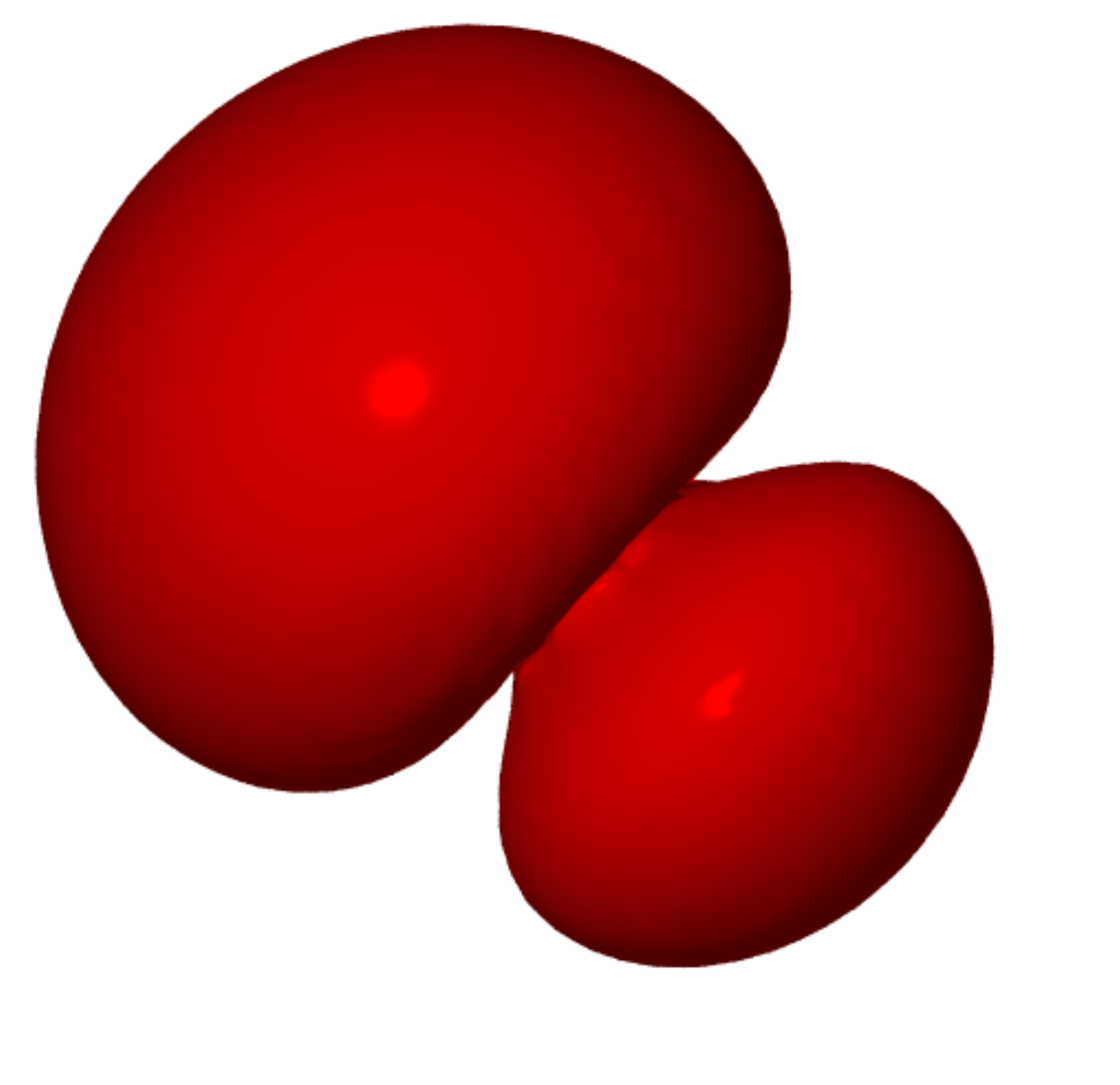}
        }
        \subfloat[N-SC$_{AvgSIC}$]{
        \includegraphics[width=0.3\linewidth]{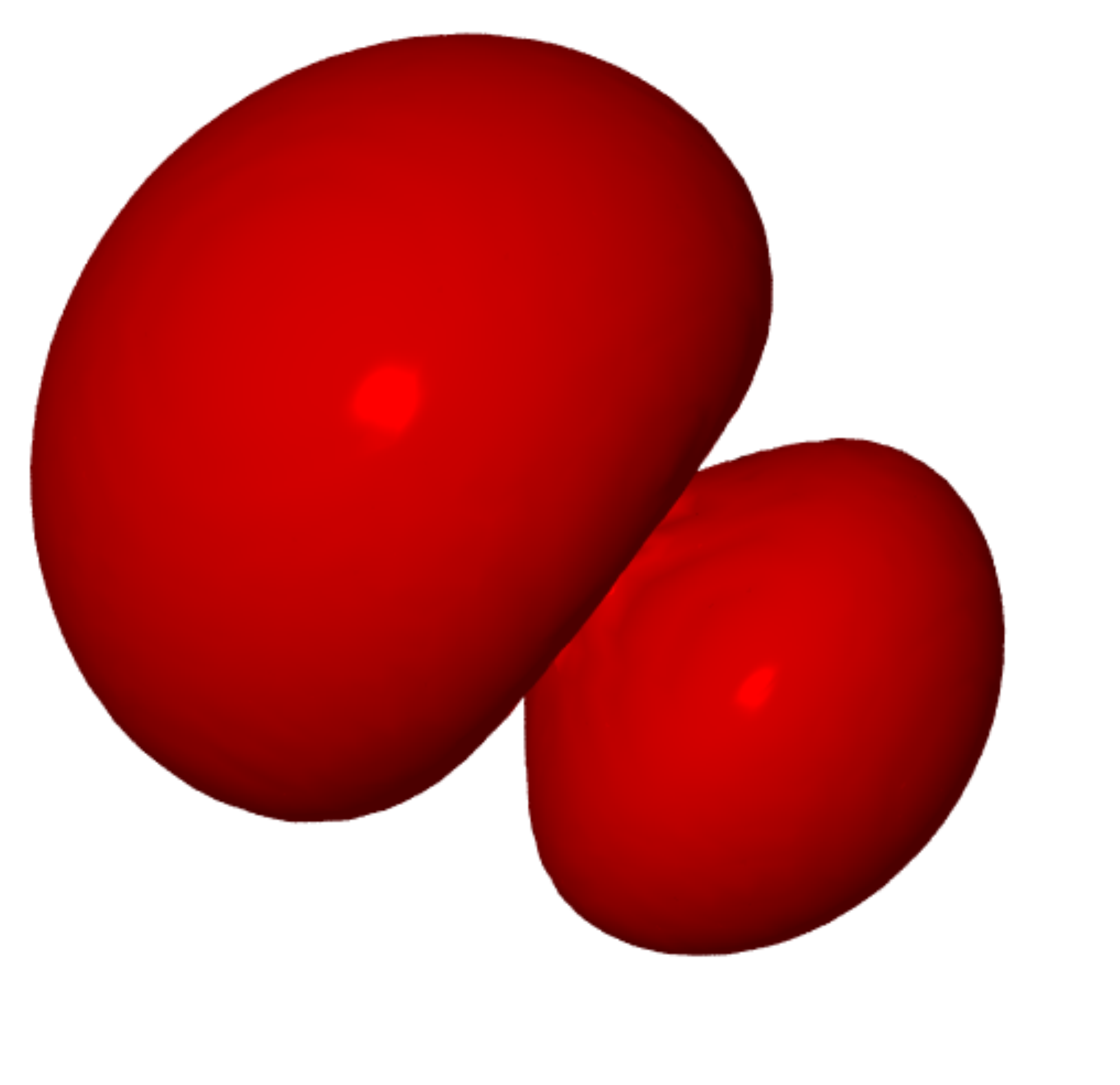}
        } \\
        \subfloat[Ne-LSDA]{
        \includegraphics[width=0.3\linewidth]{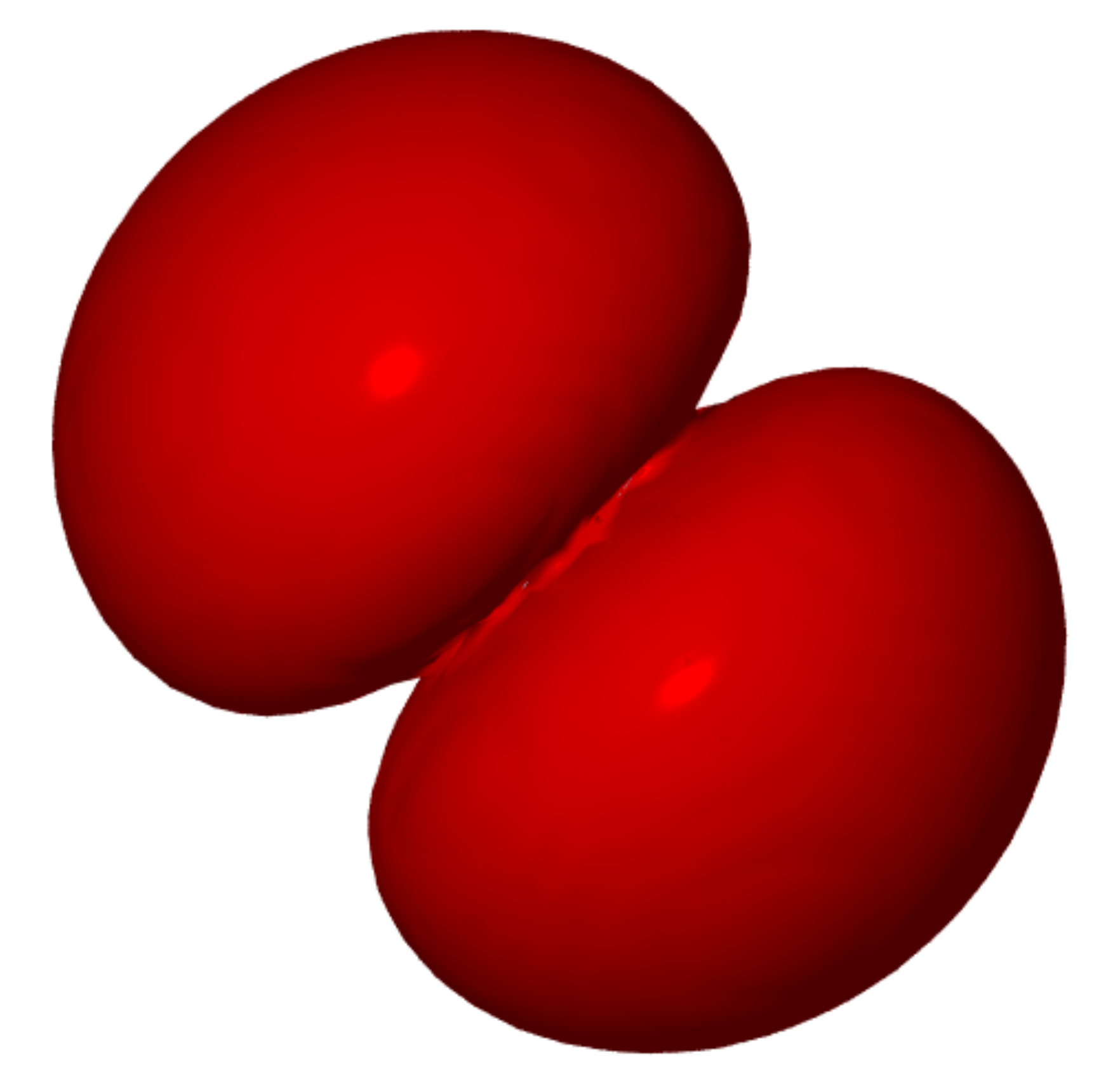}
        }
        \subfloat[Ne-SC$_{Jacobi}$]{
        \includegraphics[width=0.3\linewidth]{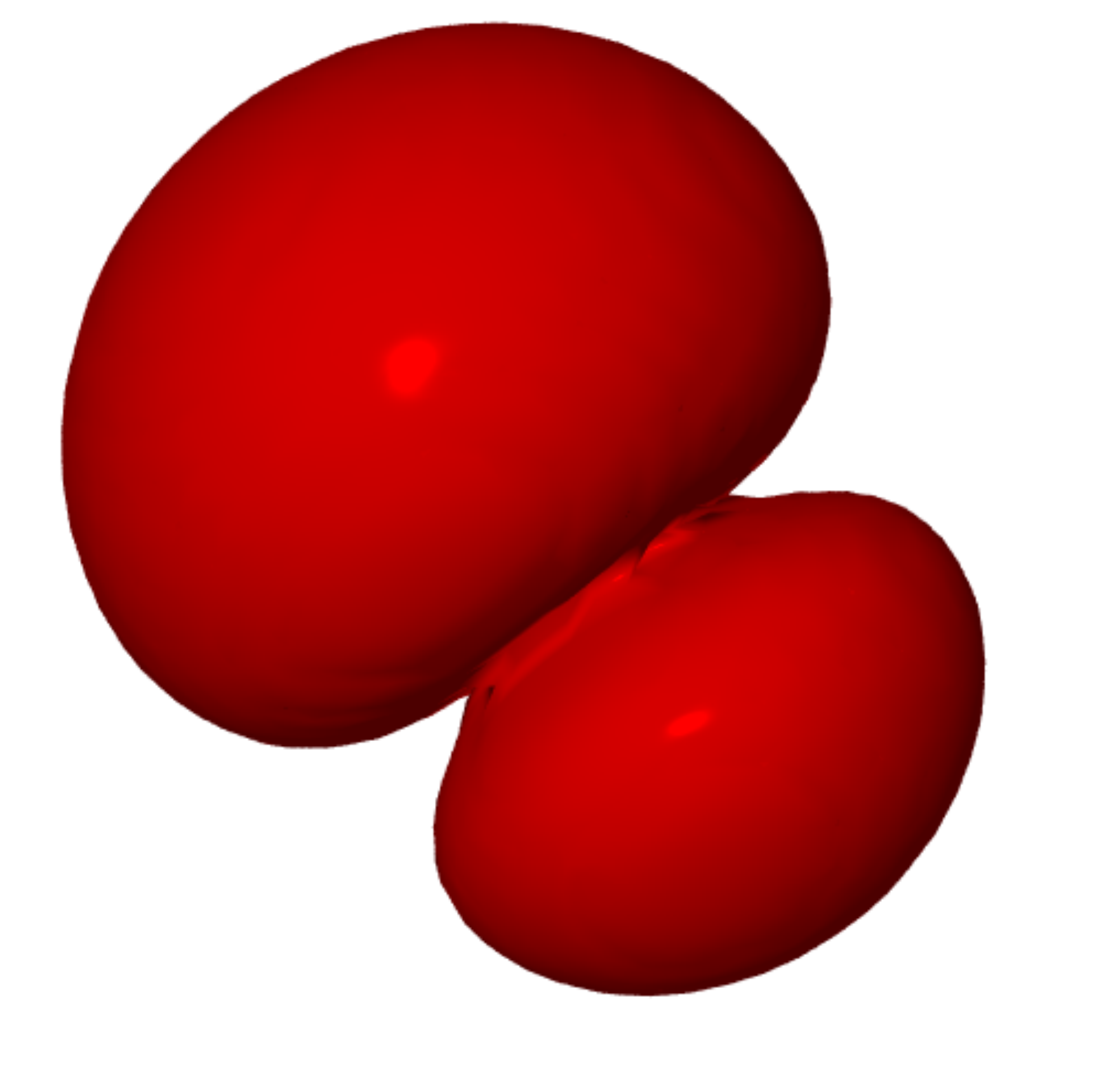}
        }
        \subfloat[Ne-SC$_{AvgSIC}$]{
        \includegraphics[width=0.3\linewidth]{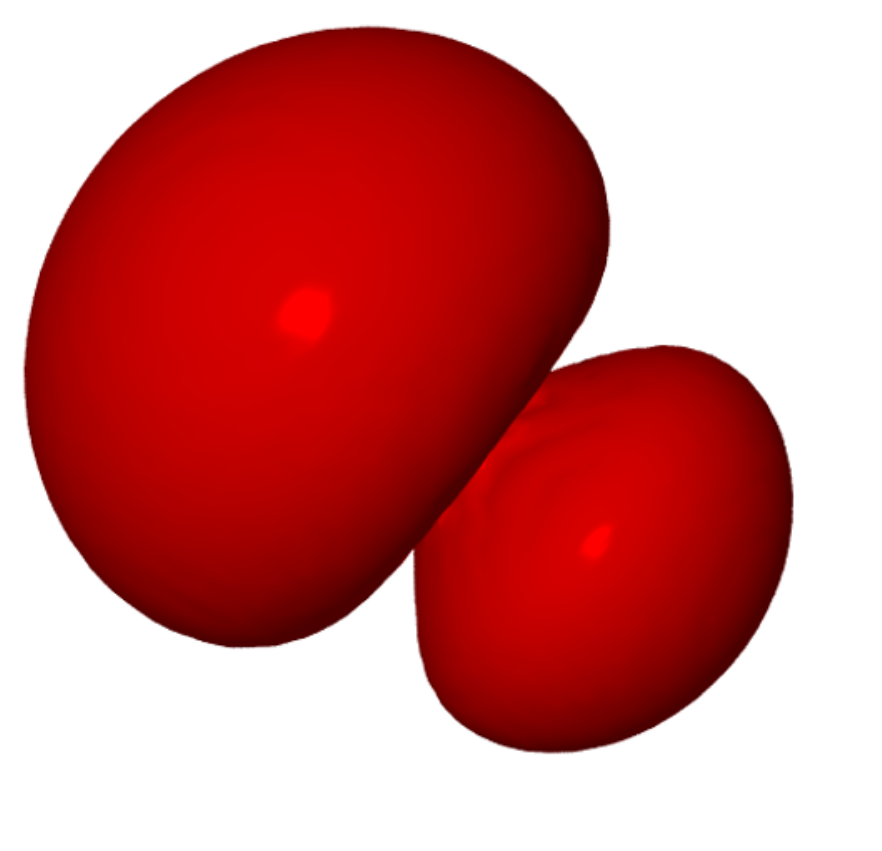}
        }
        
        \caption{The isosurface of $p$ orbitals for (a) N and (d) Ne atoms (within LSDA). Likewise, 
         the middle [(b) and (e)] and the right panel [(c) and (f)] show  isosurfaces for N and Ne atoms
        $sp$ hybridized orbitals in the  SC$_{Jacobi}$ (obtained using the FLOSIC code) and SC$_{AvgSIC}$ (present work) methods.}
        \label{fig:hybridization}
    \end{figure}

\section{Computational details}\label{sec:computational_details}
Total energies in M-SPARC calculations were converged to $10^{-3} E_h$ 
 with respect to grid spacing and domain size. 
Optimized Norm-Conserving Vanderbilt (ONCV) pseudopotentials \cite{PhysRevB.88.085117} were generated with exchange-correlation in the local density approximation (LDA) \cite{Perdew1981} based on corresponding GGA \cite{perdew1996pbe} potentials in the SG15 library. \cite{SCHLIPF201536} The ONCV formulation employs multiple nonlocal projectors in each angular momentum channel to obtain high transferability across species, configurations, and external conditions.
    All calculations were tested using LSDA exchange and the PW92 correlation functional.

    The FLOSIC code is an implementation based on the UTEP-NRLMOL code. The code uses an 
    accurate numerical integration grid scheme. \cite{Pederson1990} The default basis
    set\cite{Jackson1990} is shown to give comparable results to the cc-pVQZ basis set.\cite{schwalbe2018fermi}
    Calculations were also performed using the PyFLOSIC code\cite{Schwalbe2019}, a pySCF\cite{sun2018pyscf} based code, to produce
    benchmark calculations for comparison as it allows using basis sets with higher angular momentum functions. 
    Non-self-consistent one-shot (OS) calculations were performed by calculating SIC energies 
    at the end of a standard DFT calculation using the converged Kohn-Sham orbitals. 
    Self-consistent (SC) calculations with the M-SPARC and FLOSIC codes employed 
    the SC$_{AvgSIC}$ potential approach described in Sec.~\ref{sec:SC}. FLOSIC and 
    PyFLOSIC calculations employed SC\textsubscript{Jacobi} and SC\textsubscript{UH} approaches as well. 
    The starting FOD positions used in M-SPARC were optimized FOD positions from the FLOSIC code using 
    BHS pseudopotentials. The FOD positions were further adjusted according to the forces as needed. 
    All FODs were converged to a maximum force of 10$^{-3} E_h$/Bohr. 
    In the sections below we present results of our real-space formulation for atomic energies, 
    atomization energies, ionization potentials and barrier heights,
     with comparisons to standard Gaussian based results.

\section{Results} \label{sec:results}
\subsection{Basis set completeness} \label{Li2basis}
  The real-space methodology discretizes the Kohn-Sham equations in a chosen domain on a uniform grid in real space. For periodic systems, the computational domain is the unit cell, while for isolated systems it is a box sufficiently large to contain the wavefunction tails to desired accuracy. Thus, for atomic and molecular systems of any species and configuration, discretization errors can be reduced as far as desired by refining the grid and increasing domain size. In the present calculations, grid and domain errors were converged to 10$^{-3} E_h$ or less in all cases.
  
The systematic convergence afforded by real-space methods offers some advantages in the context of FLOSIC and other such advanced exchange-correlation formulations. In particular, since individual orbital densities vary more rapidly than the total density, higher resolution is generally required in FLOSIC calculations than for standard LDA and GGA; and this is readily obtained by refining the grid. 
In addition, where far-from-equilibrium configurations are of interest, such as for significantly stretched or compressed bonds, errors are again straightforwardly reduced as far as desired by simply refining the grid, in contrast to basis-set oriented approaches.

The FLOSIC and PyFLOSIC codes employ several basis sets to facilitate accurate calculations of a wide variety of properties.
  In addition to the default NRLMOL basis employed in the FLOSIC code,
  we also employed pc-$n$ basis sets,\cite{Jensen2001} 
  where $n=0-4$, which have been designed to systematically converge to the basis set limit as $n$ is 
  increased. At the pc-3 and pc-4 levels, the basis for lithium makes use of $f$ or higher
  angular momentum functions. As the FLOSIC code does not currently support $f$ or  
  higher angular momentum functions, the pc-3 and pc-4 results were obtained using PyFLOSIC.

\subsection{Atoms}
The hydrogen atom is the simplest system. For one electron systems, the PZ-SIC is exact.
For the hydrogen atom a single FOD is required. The SIC energy in this case is 
independent of the FOD position. We find OS-SIC calculations differ by 
0.12 m$E_h$ among all implementations. Self-consistency improves the energies to within 
0.016-0.36 m$E_h$ of the exact answer (0.5 $E_h$). By refining the grid spacing to 0.2 Bohr, the self-consistent M-SPARC 
calculation comes within 0.016 m$E_h$ of the exact energy.
Using the pc-2 basis set, the LSDA and OS-SIC calculations effectively give the same 
energy between the FLOSIC code and PyFLOSIC. Self-consistency increases the difference 
by 0.3 m$E_h$. This difference may be explained by the different implementations of 
self-consistency in each code, as detailed in Sec.~\ref{sec:SC}. 
When the SIC equations are solved using the localization equations, it was found that $s$ and $p$
orbitals are hybridized into four symmetrically equivalent orbitals. Using the Slater-averaging approach to self-consistency we find the same hybridization as shown in Fig.~\ref{fig:hybridization}.

Figure \ref{fig:hybridization}
shows the isosurfaces of $p$ orbital densities of the nitrogen and neon atom obtained with 
standard LSDA calculations along with those obtained from the FLOSIC code, which
uses Jacobi rotations to build the SIC Hamiltonian, and the present implementation,
which uses a multiplicative average SIC potential of the generalized Slater method.
These are Kohn-Sham orbitals for LSDA and FLOs for the Jacobi and the generalized Slater methods.
It is evident from the figure that the present approach, like the
FLOSIC-Jacobi scheme, also produces $sp$-hybridized orbitals.

\begin{figure}
    \centering
    \includegraphics[width=0.8\columnwidth]{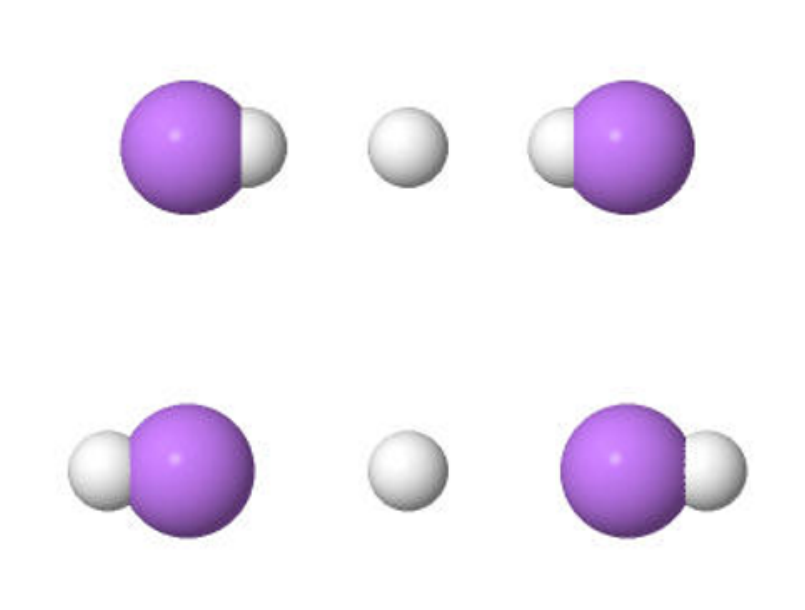}
    \caption{Two sets of FODs for Li$_2$ each at different energy minima. The Li atoms (purple) and FODs (white) are shown.}
    \label{fig:Li2}
\end{figure}

\subsection{Lithium Dimer} \label{sec:li2}
\begin{figure}[h]
    \centering
    \includegraphics[width=1.0\columnwidth]{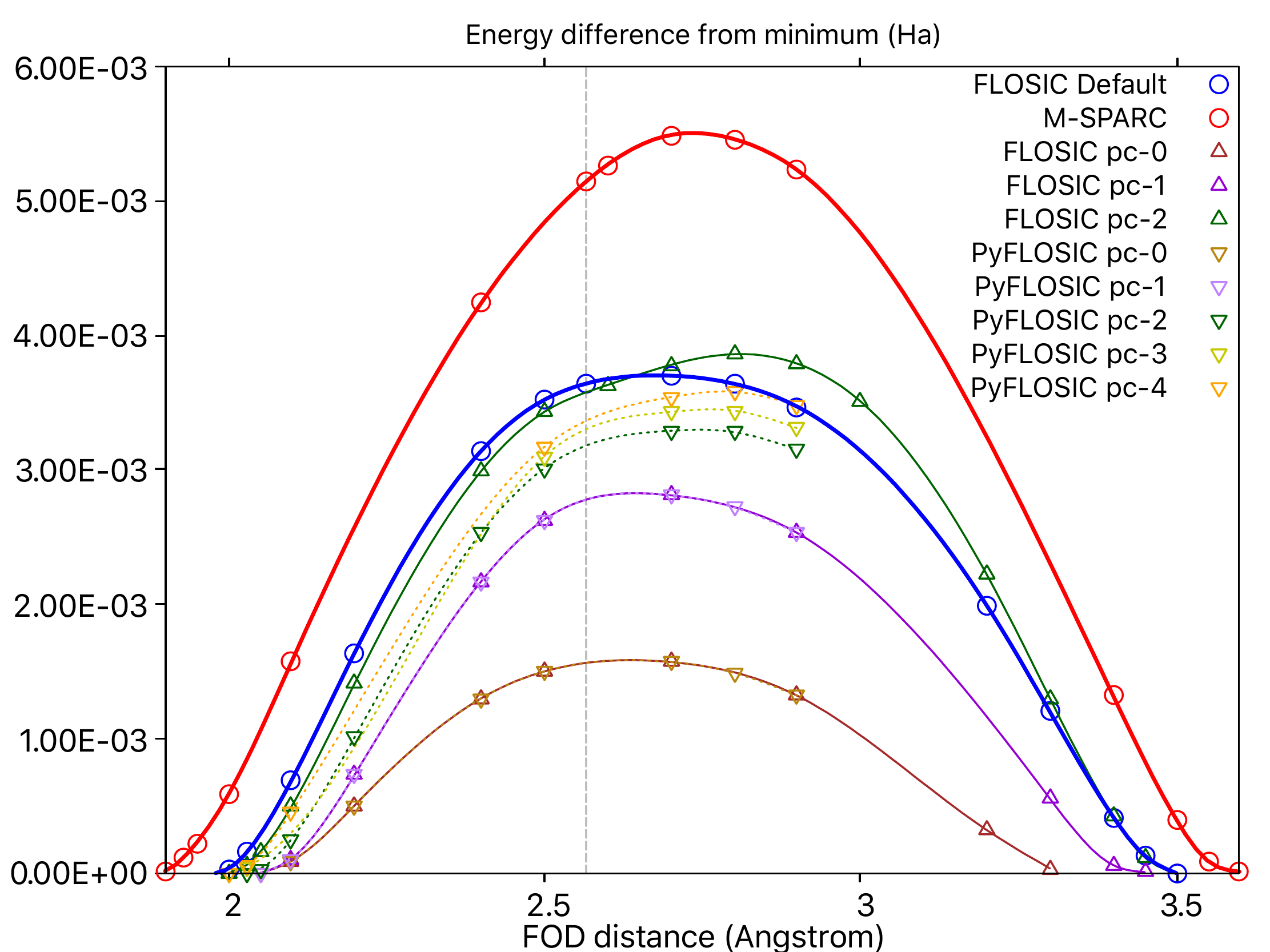}
    \caption{Energy ($E_h$) as a function of FOD distance from origin for Li$_2$. Energy plotted as difference from 
    minimum energy. PyFLOSIC calculations shown as dotted lines. Atomic position shown as dashed vertical 
    line.}
    \label{fig:basis}
\end{figure}
   We considered the Li$_2$ dimer as another test system.
   The spin-unpolarized description of Li$_2$ requires three FODs. 
   By moving the outer FODs symmetrically along the bond axis, it is possible to 
   reduce the search for optimal FOD positions to one dimension. Using this approach one
   finds two minima of equal energy, as shown in Fig.~\ref{fig:Li2}, with one corresponding to the FODs being found between the atoms and one outside the atoms. Other minima are possible with non-symmetric FOD positions, e.g. 1 outside, 1 inside.
   
   In Fig.~\ref{fig:basis} we plot the energy as a function of the FODs' distance from the origin, shifting the energy so the minimum energy is zero. The atomic position is shown as the vertical dashed line around 2.56 Bohr from the origin.
    In plotting the differences with M-SPARC and the FLOSIC code using the default basis set, we found that the FOD positions in M-SPARC were pushed farther away from the atoms for both the inside and outside minima, as well as a larger energy difference at the local maximum. 
    To determine the effect of basis set on these differences,
    we ran calculations utilizing the pc-$n$ basis sets, which approach the basis set limit as $n$ is increased from 0 to 4. We compared $n=0-2$ runs with the FLOSIC and PyFLOSIC codes to verify the implementations were consistent, and performed calculations with $n=3-4$ with PyFLOSIC.
    
    Results plotted in Fig.~\ref{fig:basis} show the FOD positions and local maximum begin approaching the M-SPARC results as the basis approaches the basis set limit, but begin to converge from $n=2-4$. 
We also tested the augmented pc-4 basis set, which adds diffuse functions, but found no difference from pc-4. 
The remaining difference in converged M-SPARC and FLOSIC/PyFLOSIC energies may be attributed to the different orbital densities in the core region, M-SPARC densities being smooth pseudo-densities and FLOSIC/PyFLOSIC being Gaussian all-electron.

\begin{figure}
    \centering
    \includegraphics[width=1.0\columnwidth]{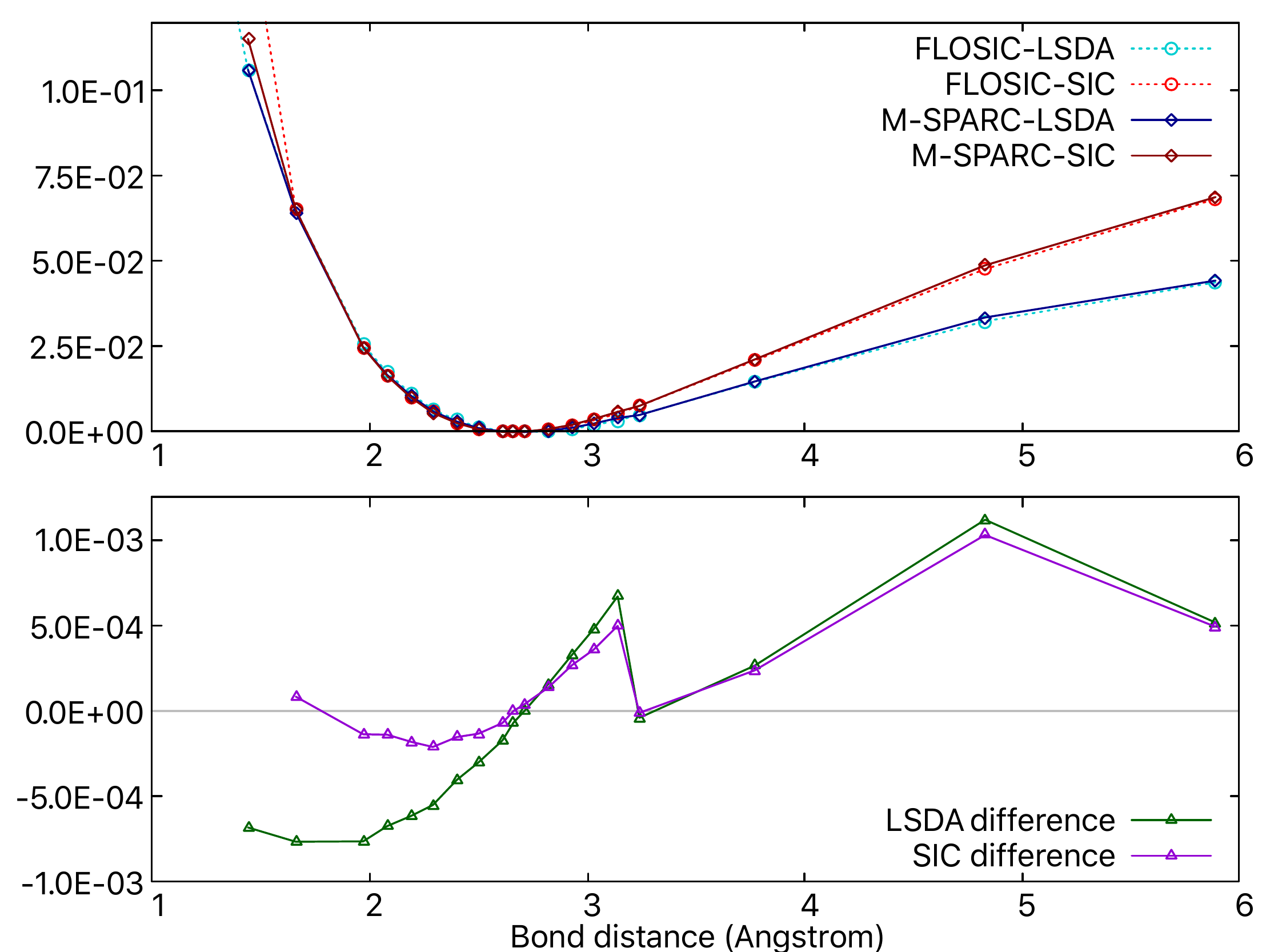}
    \caption{Top: Energy ($E_h$) as a function of bond distance for Li$_2$. Energy is shifted in each case so energy at equilibrium distance is zero. FODs for SIC calculations are optimized for each bond distance.
    Bottom: Energy difference ($E_h$) between FLOSIC and M-SPARC calculations at a given bond distance.
    }
    \label{fig:bond}
\end{figure}

While converged FOD positions may differ depending on potential representation, 
energy differences and physical observables such as equilibrium bond length should give comparable results.
    Figure~\ref{fig:bond} shows the energy vs. bond distance for the Li$_2$ molecule computed by FLOSIC and M-SPARC codes. For each bond distance, FOD forces were converged to $10^{-3}$ $E_h$/Bohr. 
    The results show excellent agreement over a wide range, as well as finding the same equilibrium bond length in both M-SPARC and FLOSIC code calculations. 
Plotting the difference between M-SPARC and FLOSIC results in Fig.~\ref{fig:bond}, we see that both implementations agree to within 
a milli-Hartree. In the M-SPARC calculations, the results must be converged with respect to domain size and 
grid spacing.
Errors will be larger as atoms approach the domain boundary. 
When the domain size is increased for the calculation around 3.2 Bohr, 
the error essentially vanishes.
At very small bond distance, optimizing the FODs leads to cases 
in which they become too close and linear dependencies arise in the L\"owdin-orthogonalization of 
the transformation matrix. 
In addition, pseudopotential overlap, fixed Gaussian exponents, and/or basis contraction may contribute to differences at the shortest bond distances.
\cite{C9CP06106A}

\begin{table*}
 \caption{  Atomization energies (in eV)  
}
 \label{table:AE}
 %\begin{tabular}{cccccc @{\extracolsep{8pt}} ccccc}
 \begin{tabular*}{\textwidth}{@{\extracolsep{\fill}}ccccccccccc}
 \hline \hline
 {}&\multicolumn{5}{c}{FLOSIC code} & \multicolumn{4}{c}{M-SPARC code} \\
 \cline{2-6} \cline{7-10}
 Molecule & LSDA &	OS-SIC &	SC$_{Jacobi}$ &	SC$_{AvgSIC}$	 & SC$_{0.5-AvgSIC}$ &	LSDA &	OS-SIC &	SC$_{AvgSIC}$	& SC$_{0.5-AvgSIC}$ &	Expt.\footnote{Reference [\onlinecite{NIST_CCCBD}]} \\
 \hline
 N$_2$ &  11.54 &  10.25  &  10.24 &   9.92  &  10.06 &  10.99  &   9.64 &   9.60  &   9.60 &   9.76  \\
 O$_2$ &   7.52 &   5.01  &   5.06 &   4.75  &   4.81 &   7.33  &   5.00 &   4.98  &   4.97 &   5.12  \\
 CO &     12.92 &  11.28  &  11.07 &  10.52  &  10.94 &  12.68  &  10.95 &  10.90  &  10.91 &  11.11  \\
 CO$_2$ &  20.40 &  15.72  &  15.88 &  14.89  &  15.34 &  20.06  &  15.46 &  15.55  &  15.49 &  16.56  \\
 C$_2$H$_2$ &  19.90 &  19.27  &  18.83 &  17.83  &  18.72 &  19.68  &  18.79 &  18.69  &  18.71 &  16.86  \\
 H$_2$  &   4.90 &   4.98  &   4.96 &   4.96  &   4.96 &   4.90  &   4.92 &   4.91  &   4.90 &   4.48  \\
 CH$_4$ &  20.03 &  20.42  &  20.20 &  19.64  &  20.10 &  19.92  &  20.23 &  20.16  &  20.14 &  17.02  \\
 NH$_3$ &  14.61 &  14.43  &  14.44 &  14.18  &  14.31 &  14.37  &  14.25 &  14.21  &  14.19 &  12.00  \\
 H$_2$O &  11.55 &  10.68  &  10.67 &  10.49  &  10.57 &  11.42  &  10.65 &  10.62  &  10.61 &   9.51  \\
 \hline
 MAE &      2.33 &  1.28  &   1.16 &   1.11  &   1.20 &   2.10  &   1.16 &   1.14  &   1.14 &     \\\hline
 MARE(\%) & 21.4 &	10.2 &	  9.3 &	   9.4 &  	 9.8 &	 19.3 &	    9.2 &	 9.0 &	   9.0 &\\
 \hline
 \end{tabular*}
\end{table*}

\subsection{Atomization Energies} \label{sec:AEs}
    The atomization energy (AE) is defined as 
   $$AE =  \sum_i^{N_{atoms}} E_i - E_{mol} > 0, $$
where $E_i$ is the energy of the constituent atom and $E_{mol}$ is the energy of the molecule. We computed AEs for a set of small molecules used in the initial FLOSIC publications \cite{Pederson2014} as well as the first self-consistent implementation. \cite{Yang2017} FOD positions were optimized within each implementation to within 10$^{-3}$ $E_h$/Bohr/FOD. All molecular geometries and initial FOD positions were taken from Ref. [\onlinecite{Yang2017}], which were obtained using the LSDA-PW92 exchange-correlation functional and the default FLOSIC code basis set.
The calculation of atomization energies requires total energies of atoms. We found that the unscaled SC$_{AvgSIC}$ approach exhibited convergence difficulties which required relaxing the FOD optimization threshold.
The lithium atom presented an additional issue in both the FLOSIC code and M-SPARC for SC$_{AvgSIC}$ calculations. During the SCF cycle for the atom, occupation of the HOO 
alternates between spin-up and spin-down orbitals from iteration to iteration.
The SIC energy of a run is determined from the spin-up orbital as specified at the beginning of the calculation, so the converged energy does not correctly reflect this spin change. Since this prevented us from comparing the SC calculations directly with the LSDA and OS calculations, molecules containing Li were left out of the reported average error.

The calculated AEs are shown in Table~\ref{table:AE} for LSDA and OS-SIC as well as self-consistent implementations described in Sec.~\ref{sec:SC}. 
We compare our results to the experimental atomization energies\cite{NIST_CCCBD} and report the mean absolute error (MAE) and mean absolute relative error (MARE) of the nine molecules.
    AEs for LSDA calculations are overestimated in both cases, but are slightly reduced in M-SPARC. The same can be seen in OS-SIC runs, where M-SPARC reduces the MARE from 10.2\% in the FLOSIC code to 9.2\% compared to experiment.

Self-consistency slightly reduces errors relative to one-shot calculations.
    M-SPARC SC$_{AvgSIC}$ results show a slight improvement of 
    $0.3-0.4$\% in the MARE compared to SC$_{Jacobi}$ and SC$_{AvgSIC}$ calculations using the FLOSIC code. 
    Self-consistent calculations with the FLOSIC code differ slightly, with SC$_{AvgSIC}$ errors slightly 
     increased on average
    compared to SC$_{Jacobi}$. 
    O$_2$, for example, increases from 1.1\% to 7.3\% relative error when using SC$_{AvgSIC}$. 
    By scaling the SIC by 0.5 with SC$_{0.5-AvgSIC}$, the error improves slightly to 6.0\%. 
    In all cases, the error is much smaller compared to the LSDA error of 46.8\%.  
    In M-SPARC, the average error remains the same for SC$_{AvgSIC}$ and scaled SC$_{0.5-AvgSIC}$ calculations, with results somewhat closer to experiment than FLOSIC code results.

\begin{table*}
 \caption{  Ionization potentials (in eV)  }
 \label{table:IP}
 %\begin{tabular}{ccccc @{\extracolsep{8pt}} cccc}
  \begin{tabular*}{\textwidth}{@{\extracolsep{\fill}}ccccccccc}
 \hline
 \hline
  {}&\multicolumn{4}{c}{FLOSIC code} & \multicolumn{3}{c}{M-SPARC code} \\
 \cline{2-5} \cline{6-8}

 Molecule &  LSDA &	SC$_{Jacobi}$ &	SC$_{AvgSIC}$ &	SC$_{0.5-AvgSIC}$ &	LSDA &	SC$_{AvgSIC}$ &	SC$_{0.5-AvgSIC}$ &	Expt.\footnote{Reference [\onlinecite{Zhang2007}]} \\
 \hline
 N$_2$ & 10.43 & 17.79  & 18.16 & 14.22  & 10.44 & 17.75  & 14.05 &  15.58  \\
 O$_2$ &  7.05 & 15.95  & 16.68 & 11.74  &  7.14 & 16.47  & 11.73 &  12.30  \\
 CO &     9.15 & 15.70  & 16.20 & 12.60  &  9.16 & 15.85  & 12.46 &  14.01  \\
 CO$_2$ &  9.31 & 16.07  & 17.04 & 13.11  &  9.33 & 16.73  & 12.99 &  13.78  \\
 C$_2$H$_2$ &  7.34 & 12.59  & 13.48 & 10.35  &  7.34 & 13.29  & 10.28 &  11.49  \\
 LiF &     6.26 & 13.77  & 14.44 & 10.19  &  6.32 & 14.42  & 10.26 &  11.30  \\
 H$_2$ &  10.22 & 16.78  & 16.80 & 13.47  & 10.03 & 16.45  & 13.20 &  15.43  \\
 Li$_2$ &  3.23 &  5.48  &  5.89 &  4.51  &  3.16 &  5.84  &  4.45 &   5.11  \\
 CH$_4$ &  9.43 & 15.90  & 16.00 & 12.67  &  9.44 & 16.04  & 12.70 &  13.60  \\
 NH$_3$ &  6.27 & 12.61  & 12.96 &  9.54  &  6.28 & 12.90  &  9.53 &  10.82  \\
 H$_2$O &  7.33 & 14.87  & 14.99 & 11.09  &  7.37 & 14.95  & 11.09 &  12.62  \\
 \hline
 MAE &        4.55 &   1.95  &   2.42 &   1.14  &   4.55 &   2.24  &   1.21 &     \\
 \hline
 MARE(\%) &  37.1 &   15.6  &   19.7 &    9.4  &    37.2 &   18.4  &   9.9 &     \\
 \hline
 \end{tabular*}
\end{table*}

\subsection{Ionization potentials}
The Kohn-Sham eigenvalues are not electron removal energies except for the highest occupied orbital (HOO).
In exact DFT, the HOO eigenvalue equals the negative of the ionization 
potential.\cite{perdew1982density,levy1984exact,almbladh1985exact,perdew1997comment,PhysRevB.60.4545} 
This relationship  does not strictly hold for approximate density functionals used in 
practical applications of KS-DFT and in most DFAs the absolute value of the 
HOO eigenvalue substantially underestimates the first ionization potential due to SIE
of the exchange-correlation potential. This can be seen easily. 
In this approximation the asymptotic 
decay of the electron density is primarily governed by the HOO density. Thus, 
in this limit, the weight factor $\tilde{\rho}_i/\rho$ approaches unity. Therefore, 
correcting for self-interaction results in the $-1/r$ asymptotic behavior
of the exchange-correlation potential and provides a significant improvement
in the HOO eigenvalue.

In Table~\ref{table:IP}, we show the HOO eigenvalues obtained by the present real-space M-SPARC and
existing Gaussian-based FLOSIC codes. 
We compare the computed HOO eigenvalues against the experimental ionization potentials (IPs)\cite{Zhang2007} and report MAEs and MAREs for the set.
The LSDA M-SPARC and FLOSIC code eigenvalues agree well, with MAE of 4.55 eV relative to experiment.
The application of SIC using the SC$_{AvgSIC}$ method reduces the MAE from 4.55 eV to 2.42 eV
for the FLOSIC code and from 4.55 eV to 2.24 eV for M-SPARC. 
Although the HOO eigenvalues in the
SC$_{AvgSIC}$ method are improved with respect to LSDA, they are overestimated 
with respect to experimental eigenvalues. This overestimation can be mitigated 
by scaling down the $\bar{v}_{SIC}^{\sigma}$ potential. The HOO eigenvalues obtained by scaling
down the $\bar{v}_{SIC}^{\sigma}$ potential by 0.5 show clear improvements with further
decreases in MAE to 1.14 eV for the FLOSIC code and to 1.21 eV for M-SPARC.

\subsection{Barrier heights of chemical reactions}
Finally, we test the present implementation using the BH6 database of barrier heights in chemical reactions.  
BH6 is a representative subset of the BH24 set. \cite{doi:10.1021/ct600281g} The reactions included in BH6 are:
OH + CH$_4 \rightarrow$ CH$_3$ + H$_2$O, H + OH $\rightarrow$ H$_2$ + O, and H + H$_2$S $\rightarrow$ H$_2$ + HS. 
Total energies at the left hand side, the right hand side, and the saddle 
point of these chemical reactions were calculated, and the barrier heights of 
the forward (f) and reverse (r) reactions were obtained by taking the relevant energy differences. 
The geometries used are same as those provided in Ref. [\onlinecite{doi:10.1021/ct600281g}].
The reference values are from Ref. [\onlinecite{doi:10.1021/jp035287b}].
The results for the methods considered are summarized in Table \ref{table:bh6}.
The local (and also semi-local)  DFAs 
underestimate barrier heights \cite{doi:10.1063/1.2176608} 
by giving transition state energies that are too low compared to the reactant and 
product energies.  Accurate description of transition states 
require  full nonlocality in the exchange-correlation potential that  
local (and semi-local) approximations lack. 
The present SC$_{AvgSIC}$ scheme  
further improves over SC$_{Jacobi}$.
The MAE in barrier heights in the  
 SC$_{AvgSIC}$ implementation is 2.8 kcal/mol vs. 4.9 kcal/mol for the SC$_{Jacobi}$ method,
 corresponding to a 14.1\% improvement in MARE.

\begin{table*}
    \caption{Barrier heights in kcal/mol of the forward (f) and reverse (r) chemical reactions belonging to the BH6 database. Signed errors are shown.}
    \label{table:bh6}
    %\begin{tabular}{cccc @{\extracolsep{8pt}} cccc}
     \begin{tabular*}{\textwidth}{@{\extracolsep{\fill}}cccccccc}
    \hline \hline
     {}& & \multicolumn{2}{c}{FLOSIC code} & \multicolumn{3}{c}{M-SPARC code} &\\
    \cline{3-4} \cline{5-7}
     Reaction & Barrier &	LSDA &	SC$_{Jacobi}$ \footnote{Reference [\onlinecite{doi:10.1063/1.5129533}]} &		 LSDA &	OS-SIC &	SC$_{AvgSIC}$	& 	Ref.  \footnote{Reference [\onlinecite{doi:10.1021/jp035287b}]} \\
\hline

     OH + CH$_4$ $\rightarrow$ CH$_3$ + H$_2$O  
        &  f  &  -23.6  &       -2.2  &      -22.9  &  1.0   &   1.0 &   6.7\\
        &  r  &  -17.4  &      -12.5  &      -17.0  &  -6.5  &  -6.6 &  19.6\\
    H + OH $\rightarrow$ H$_2$ + O  
        &  f  &  -11.8  &      -1.1  &       -13.1  &  -1.6  &  -0.1 &  10.7\\
        &  r  &  -25.3  &      -4.8  &       -24.8  &  -0.5  &   0.9 &  13.1\\
    H + H$_2$S $\rightarrow$ H$_2$ + HS  
        &  f  &  -10.3  &      -1.7  &       -10.0  &  -0.9  &  -1.0 &   3.6\\
        &  r  &  -17.2  &      -7.0  &       -16.5  &  -3.9  &  -6.0 &  17.3\\
    \hline
    MAE  &    &  17.6  &        4.9  &        17.4  &  2.4   &  2.8 & \\
    \hline
    MARE  &   &  188.3\%  &    38.5\%  &   185.7\%  &  19.0\%  &  24.4\% & \\
    \hline
        
    \end{tabular*}
\end{table*}

\section{Conclusion}
We presented a formulation and implementation of the FLOSIC self-interaction correction in real space.
The real-space formulation allows rigorous, systematic convergence 
for all atomic species and configurations. 
The present M-SPARC based implementation demonstrated the feasibility and accuracy of the real-space formulation.
Self-consistency in the present work was introduced using a generalized 
Slater statistical average SIC potential. 

We verified our M-SPARC based real-space implementation 
by computing atomization energies and ionization potentials of selected molecules and comparing to those obtained by the established Gaussian-based FLOSIC and PyFLOSIC codes.
We compared results for LSDA and non-self-consistent as well as
self-consistent FLOSIC calculations. The results obtained with the M-SPARC code show 
good agreement with those obtained by the Gaussian-based FLOSIC code,
with M-SPARC results somewhat closer to experiment for the systems considered. 
The application of SIC with Slater-averaged SIC potentials showed improvement in HOOs over LSDA. 
We found that scaling down the SIC potential by a factor of 0.5 
gives comparable or improved results compared to the  all-electron Gaussian-based 
FLOSIC code implementation. 
The MAE in computed ionization potentials  
with respect to experiment was found to be 2.24 eV and decreases to 1.21 eV 
when the SIC potential is scaled down by 0.5. 
We also found that the barrier heights 
obtained using the present SC$_{AvgSIC}$ scheme are further improved over the SC$_{Jacobi}$ 
barrier heights.   
Our initial results indicate that the real-space formulation
provides an accurate and systematically improvable approach for FLOSIC calculations in the Slater-average form.
Future work will include
implementation in the large-scale parallel SPARC code
and extension to a full OEP formulation.

\section{ Acknowledgements}
C.M.D., T. B., and R. R. Z. acknowledge discussions with Dr. Yoh Yamamoto and thank 
Prof. Koblar Jackson  for reading the manuscript.
This material is based upon work supported by the U.S. Department of Energy, Office of Science,
Office of Workforce Development for Teachers and Scientists, Office of Science Graduate Student Research
(SCGSR) program. The SCGSR program is administered by the Oak Ridge Institute for Science and Education
(ORISE) for the DOE. ORISE is managed by ORAU under contract number DE-SC0014664. 
C.M.D., T.B. and R.R.Z. acknowledge support by the US Department of Energy, Office of
Science, Office of Basic Energy Sciences, as part of the Computational Chemical Sciences
Program under Award No. DE-SC0018331. 
This work was performed in part under the auspices of the U.S. Department of Energy by Lawrence Livermore National Laboratory under Contract DE-AC52-07NA27344.
P.S. and Q.X. acknowledge support by the grant DE-SC0019410 funded by the U.S. Department of Energy, Office of Science. All opinions expressed
in this paper are the authors' and do not necessarily reflect the policies and views of DOE, ORAU, or ORISE.

\section{Data availability}
The data that support the findings of this study are available from the corresponding author upon reasonable request.

\section{References}
\bibliography{realspaceflosic}

\end{document}